\documentclass[prb,showpacs,byrevtex,floatfix,twocolumn]{revtex4}
\usepackage{graphicx}
\usepackage{dcolumn}
\usepackage{bm}

\begin{document}

\bibliographystyle{apsrev}

%\preprint{Draft - not for distribution}

\title[inverse]{Extracting the electron--boson spectral function
$\alpha^2$F($\omega$) from infrared and photoemission data using
inverse theory}

\author{S. V. Dordevic}
\email{sasa@bnl.gov}%
\author{C. C. Homes}
\author{J.J.~Tu}
\author{T.~Valla}
\author{M.~Strongin}
\author{P.D.~Johnson}
\author{G.D.~Gu}

\affiliation{Department of Physics, Brookhaven National Laboratory,
  Upton, New York 11973}
\author{D. N. Basov}
\affiliation{Department of Physics, University of California, San
Diego, La Jolla, CA 92093}

\date{\today}

%
% The abstract goes here
%
\begin{abstract}
We present a new method of extracting electron-boson spectral
function $\alpha^2$F($\omega$) from infrared and photoemission
data. This procedure is based on inverse theory and will be shown
to be superior to previous techniques. Numerical implementation of
the algorithm is presented in detail and then used to accurately
determine the doping and temperature dependence of the spectral
function in several families of high-T$_c$ superconductors.
Principal limitations of extracting $\alpha^2$F($\omega$) from
experimental data will be pointed out. We directly compare the IR
and ARPES $\alpha^2$F($\omega$) and discuss the resonance
structure in the spectra in terms of existing theoretical models.
\end{abstract}

%
% PACS numbers
%
% 74.25.-q - General properties; correlations between normal and SC states
% 74.25.Gz - optical properties of superconductors
% 78.30.-j - Infrared and Raman spectra
%
\pacs{74.25.-q, 74.25.Gz, 78.30.-j}

\maketitle

\makeatletter%
\global\@specialpagefalse%
\let\@evenhead\@oddhead%
\makeatother

\section{Introduction}
\label{introduction}

The electron-boson spectral function is one of the most important
properties of a BCS superconductor \cite{carbotte90}. In
conventional superconductors the electron-phonon spectral function
has been successfully obtained using tunneling \cite{mcmillian65}
and infrared (IR) spectroscopy
\cite{timusk74,timusk76,marsiglio98}. The situation is more
complicated in cuprates where the mechanism of superconductivity
is still a matter of debate. Based on IR data it was suggested
very early that charge carriers in cuprates might be strongly
coupled to some collective boson mode \cite{thomas88}. It was
subsequently proposed that this collective mode might be magnetic
in origin \cite{pines90,carbotte99,munzar99}. Within this scenario
electrons are strongly coupled to a so called ``41\,meV" resonance
peak observed in INS (Refs.~\onlinecite{bourges99,mook01}). The
peak is believed to originate from antiferromagnetic spin
fluctuations that persist into the superconducting state; coupling
of electrons to this mode in turn leads to Cooper pairing. However
recently this view was challenged by a proposal that charge
carriers might be strongly coupled to phonons
\cite{bogdanov00,lanzara01,shen03,zhou03}. This controversial
suggestion has revitalized the debate about whether a collective
boson mode is responsible for superconductivity in the cuprates.
An accurate and reliable determination of the electron--boson
spectral function has become essential.

In this paper we propose a new way of extracting the spectral
function from IR and Angular Resolved Photoemission Spectroscopy
(ARPES) data. The proposed method is based on inverse theory
\cite{inverse-comm}, and will be shown to have numerous advantages
over previously employed procedures. An advantage of the method is
that it eliminates the need for differentiation of the data, that
was previously the most serious problem. The inversion algorithm
uncovers extreme sensitivity of the solution to smoothing, and
offers a smoothing procedure which eliminates arbitrariness. Since
the spectral function is convoluted in the experimental data, some
information is inevitably lost; we will use inverse theory to set
the limits on useful information that can be extracted from the
data. Unlike previous techniques which are valid only at T=\,0\,K,
the new method can be applied at {\it any} temperature.

The paper is organized as follows. First in
Section~\ref{numerical} we outline the numerical procedure of
solving integral equations. In Section~\ref{example} we
demonstrate the usefulness of the new method by applying it to
previously published data for YBa$_2$Cu$_3$O$_{7-\delta}$ (Y123).
In Section~\ref{model} model calculations of spectral function
will unveil some important problems encountered when solving
integral equations. Section~\ref{negative} discusses the origin of
negative values in the spectral function and methods for dealing
with them. In Section~\ref{sc} the effect superconducting energy
gap has on the spectral function will be analyzed. In
Section~\ref{bi2212} we study the temperature dependence of the
spectral function for optimally doped
Bi$_2$Sr$_2$CaCu$_2$O$_{8+\delta}$ (Bi2212). In
Section~\ref{arpes} inverse theory is applied to ARPES data and
the spectral function of molybdenum surface Mo(110) and Bi2212
have been studied. Finally, Section~\ref{comparison} contains
quantitative comparison of the spectral functions of optimally
doped Bi2212, extracted from both IR and ARPES data; the observed
results are critically compared against existing theoretical
models. In Section~\ref{conclusions} we summarize all the major
results.

\section{Numerical procedure}
\label{numerical}

The (optical) scattering rate of electrons in the presence of
electron-phonon coupling at T=\,0\,K is given by famous Allen's
result \cite{allen71}:

\begin{equation}
\frac{1}{\tau(\omega)}=\frac{2\pi}{\omega}
\int_0^{\omega}d\Omega(\omega-\Omega)\alpha^2F(\Omega),
\label{eq:tau1}
\end{equation}
where $\alpha^2$F($\omega$) is the electron-phonon spectral
function. The scattering rate 1/$\tau(\omega)$ can be obtained
from complex optical conductivity
$\sigma(\omega)=\sigma_1(\omega)+i\sigma_2(\omega)$:

\begin{equation}
\frac{1}{\tau(\omega)}=\frac{\omega_{p}^{2}}{4 \pi}
\frac{\sigma_1(\omega)} {\sigma_1^2(\omega)+\sigma_2^2(\omega)},
\label{eq:tau}
\end{equation}
where $\omega_p$ is the conventional plasma frequency. Recently
Marsiglio, Startseva and Carbotte \cite{marsiglio98} have defined
a function W($\omega$):

\begin{equation}
W(\omega)=\frac{1}{2\pi}\frac{d^2}{d\omega^2} \left[\omega \cdot
\frac{1}{\tau(\omega )} \right], \label{eq:w1}
\end{equation}
which they claim to be W($\omega$)$\approx$\,$\alpha^2$F($\omega$)
in the phonon region \cite{marsiglio98}. It is easy to show (by
substitution, for example) that Allen's formula
Eq.~(\ref{eq:tau1}) and Eq.~(\ref{eq:w1}) are equivalent
expressions, provided 1/$\tau$($\omega$\,=\,0)=\,0.
Eq.~(\ref{eq:w1}) is frequently used to extract the spectral
function in the cuprates from IR data
\cite{carbotte99,schachinger00,schachinger01,wang01,abanov01,singley01,tu02}.
Obviously this method introduces significant numerical difficulty
since the {\it second} derivative of the data is needed. The
experimental data must be (ambiguously) smoothed ``by hand" before
Eq.~(\ref{eq:w1}) can be applied, otherwise the noise will be
amplified by (double) differentiation and will completely dominate
the solution. An alternative approach is to fit the scattering
rate with polynomials and then perform differentiation
analytically \cite{tu02,wang03}. Note also that although
Eq.~(\ref{eq:w1}) is valid only at T=\,0\,K, it is frequently
applied to higher T, even at room temperature.

Here we propose a new method of extracting the spectral function.
It is based on the following formula for the scattering rate at
{\it finite} temperatures derived by Shulga {\it et al.}
\cite{shulga91,comm-allen}:

\begin{eqnarray}
&&\frac{1}{\tau(\omega,T)}=\frac{\pi}{\omega}
\int_0^{\infty}d\Omega\alpha^2F(\Omega,T) \Big[2\omega
\coth\Big(\frac{\Omega}{2T}\Big)- \nonumber \\ &&
(\omega+\Omega)\coth\Big(\frac{\omega+\Omega}{2T}\Big)+(\omega-\Omega)
\coth\Big(\frac{\omega-\Omega}{2T}\Big)\Big],\nonumber\\
\label{eq:tau2}
\end{eqnarray}
which in the limit T$\rightarrow$\,0\,K reduces to Allen's result
Eq.~(\ref{eq:tau1}) (Ref.~\onlinecite{puchkov96rev}). Unlike
Eq.~(\ref{eq:tau1}) which has a differential form
Eq.~(\ref{eq:w1}), there is no such simple expression for
Eq.~(\ref{eq:tau2}). Therefore in order to obtain
$\alpha^2$F($\omega$) from Eq.~(\ref{eq:tau2}) one must apply
inverse theory \cite{nr}. Like most inverse problems, obtaining
the spectral function from the scattering rate data is an
ill-posed problem which requires special numerical treatment. The
spectral function appears under the integral, an operator which
has smoothing properties. That means that some of the information
on $\alpha^2$F($\omega$) is inevitably lost. Using inverse theory
our goal will be to extract as much of useful information as we
can, and set the limits on lost information.

Numerically the procedure of solving an integral equation reduces
to an optimization problem, i.e. finding the ``best" out of all
possible solutions \cite{nr}. Different criteria can be adopted
for the ``best" solution, such as: 1) closeness to the data in the
least square sense (we will call this solution ``exact") or 2)
smoothness of the solution. The most useful solution is often a
trade-off between these two.

Eq.~(\ref{eq:tau2}) is a Fredholm integral equation of the first
kind \cite{nr}; it may be rewritten as:

\begin{equation}
\frac{1}{\tau(\omega, T)}=\int_0^{\infty}d\Omega
\alpha^2F(\Omega, T) K(\omega, \Omega, T) \label{eq:tau3}
\end{equation}
where 1/$\tau(\omega,T)$ is experimental data (from
Eq.~(\ref{eq:tau})), K($\omega$,$\Omega$, T) (contains the
prefactor $\pi/\omega$ from Eq.~(\ref{eq:tau2})) is a so-called
kernel of integral equation, and $\alpha^2$F($\omega$, T) is the
unknown function to be determined. When discretized in both
$\omega$ and $\Omega$ Eq.~(\ref{eq:tau3}) becomes:

\begin{equation}
\frac{1}{\tau(\omega_i, T)}=\sum_{j=1}^{N} \Delta\Omega_j
\alpha^2F(\Omega_j, T) K(\omega_i, \Omega_j, T), \label{eq:tau4}
\end{equation}
with $i=1,N$. In matrix form:

\begin{equation}
\vec{\gamma}=K\vec{a}, \label{eq:m1}
\end{equation}
where vector $\vec{\gamma}$ corresponds to
1/$\tau$($\omega_i$,\,T), vector $\vec{a}$ to
$\alpha^2$F($\Omega_j$,\,T) and matrix K to K($\omega_i$,
$\Omega_j$, T) (Ref.~\onlinecite{quadratic-comm}). The problem is
reduced to finding vector $\vec{a}$, i.e. the inverse of matrix K.
To perform this matrix inversion we adopt a so called {\it
singular value decomposition} (SVD) \cite{nr} because it allows a
physical insight into the inversion process and offers a natural
way of smoothing. Matrix $K$ is decomposed into the following
form:

\begin{equation}
K=U [diag(w_j)] V^T \label{eq:svd1}
\end{equation}
where U and V are orthogonal matrices (U$^T$=U$^{-1}$ and
V$^T$=V$^{-1}$), and $diag(w_j)$ is a diagonal matrix with
elements $w_j$. The inverse of K is now trivial: K$^{-1}$=V
[$diag(1/w_j)$]U$^T$ and the solution to Eq.~(\ref{eq:m1}) is than
simply:

\begin{equation}
\vec{a}=K^{-1}\vec{\gamma}=V [diag(1/w_j)]U^T \vec{\gamma}.
\label{eq:m2}
\end{equation}

The elements of diagonal matrix $w_j$ are called {\it singular
values} (s.v.); they are by definition positive and are usually
arranged in decreasing order. If all of them are kept in
Eq.~(\ref{eq:m2}) the ``exact" solution, i.e. the best agreement
with the original data, is obtained. If needed, and it almost
always is when solving integral equations, the smoothing of the
solution (not the experimental data) is achieved by replacing the
largest 1/$w_j$ in Eq.~(\ref{eq:m2}) with zeros, before performing
matrix multiplications. This is a common procedure of filtering
out high frequency components in the solution \cite{nr}.

\section{An example}
\label{example}

To demonstrate the usefulness of this procedure we first analyze
the existing IR data for underdoped YBa$_2$Cu$_3$O$_{6.6}$ with
T$_c$=\,59\,K (Ref.~\onlinecite{carbotte99}).
Spectral function W($\omega$) for
this compound was previously determined using Eq.~(\ref{eq:w1}),
after 1/$\tau(\omega,T)$ had been (heavily) smoothed
\cite{carbotte99}. Here we apply the numerical procedure described
in the previous section on the same data set. We start with
1/$\tau(\omega,T=10K)$ data in the range 10-3,000 cm$^{-1}$, and
form a linear set of 300 equations to be solved (N=\,300 in
Eq.~(\ref{eq:tau4})), i.e. 300-element vectors $\vec{a}$ and
$\vec{\gamma}$ and a 300$\times$300 matrix K (Eq.~(\ref{eq:m1})).
We then decompose matrix K (Eq.~(\ref{eq:svd1})) and choose how
many of its singular values we are going to keep. Finally we
invert the matrix and solve the system for vector $\vec{a}$
(Eq.~(\ref{eq:m2})), i.e. $\alpha^2$F($\omega$) in the range
between 10-3,000 cm$^{-1}$, at 300 points.

\begin{figure}[t]
\vspace*{+0.0cm}%
\centerline{\includegraphics[width=9cm]{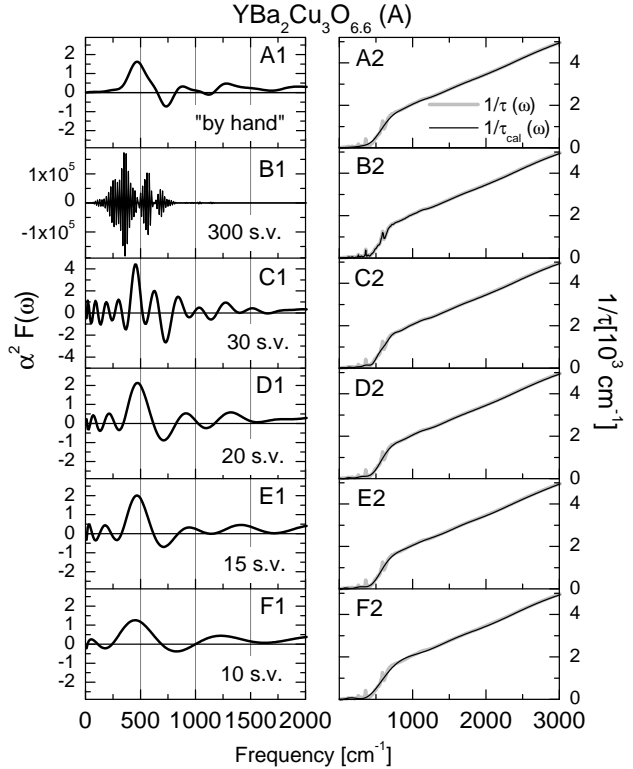}}%
\vspace*{-0.5cm}%
\caption{Spectral function $\alpha^2$F($\omega$) for underdoped
YBa$_2$Cu$_3$O$_{6.6}$ with T$_c$=\,59\,K. The left panels show
$\alpha^2$F($\omega$) and the right panels the experimental
1/$\tau(\omega)$ along with 1/$\tau_{cal}(\omega)$ calculated from
the corresponding spectral function (Eq.~(\ref{eq:tau2})). The top
panel shows previously published spectral function
\protect\cite{carbotte99} obtained from the scattering rate
smoothed ``by hand". The other five pairs of panels are the data
obtained using inverse theory. Different number of singular values
are kept in the calculations, which results in different levels of
smoothing. Note that the vertical scale in panels B1 and C1 is
different.} \label{fig:num1}
\end{figure}

\begin{figure}[t]
\vspace*{+2.0cm}%
\centerline{\includegraphics[width=9cm]{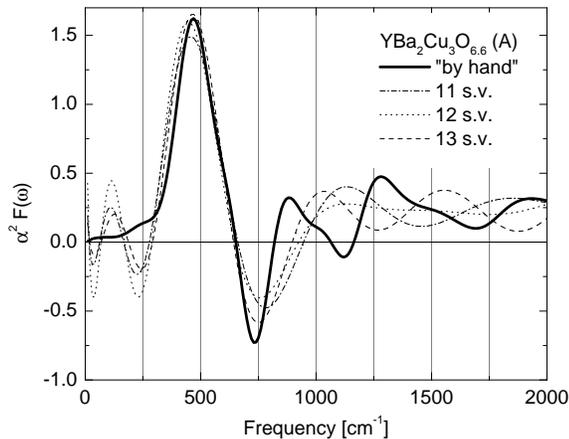}}%
\vspace*{-2.0cm}%
\caption{Spectral function $\alpha^2$F($\omega$) for underdoped
YBa$_2$Cu$_3$O$_{6.6}$ with T$_c$=\,59\,K obtained by smoothing
the experimental 1/$\tau(\omega)$ data ``by hand"
\protect\cite{carbotte99} and with SVD with 11, 12 and 13 s.v.}
\label{fig:num2}
\end{figure}

Left panels of Fig.~\ref{fig:num1} show results of
$\alpha^2$F($\omega$) calculations for YBa$_2$Cu$_3$O$_{6.6}$ at
10 K, for 6 different levels and/or methods of smoothing. The
right panels show the scattering rate 1/$\tau(\omega)$, along with
the calculated scattering rate 1/$\tau_{cal}(\omega)$, obtained by
substituting the corresponding $\alpha^2$F($\omega$) on the left
back into Eq.~(\ref{eq:tau2}). The top panels (A1 and A2) display
previously published solution \cite{carbotte99} obtained using
Eq.~(\ref{eq:w1}) after the data had been smoothed ``by hand". The
next two panels (B1 and B2) present the ``exact" solution using
SVD, with all 300 singular values different from zero. This
solution does not appear to be very useful (note the vertical
scale), although it gives the best agreement between the
experimental data 1/$\tau(\omega)$ and calculated scattering rate
1/$\tau_{cal}(\omega)$ (panel B2). One might say that the solution
contains too much information, as it unnecessarily reproduces all
the fine details in the original 1/$\tau(\omega)$ data, including
the noise. The remaining panels show SVD calculations with 30 (C),
20 (D), 15 (E) and 10 (F) biggest s.v. different from zero.
Surprisingly only a few singular values (less that 10\,$\%$ of the
total number) are needed to achieve similar spectral function as
obtained previously by smoothing the data ``by hand" (panel A1).
Indeed Fig.~\ref{fig:num2} shows that approximately 12 or 13
non-zero singular values are needed. Note however that neither of
the curves matches exactly the curve obtained from the data
smoothed ``by hand".

The $\alpha^2$F($\omega$) spectra (Fig.~\ref{fig:num2}) display
characteristic shape with a strong peak at 480\,cm$^{-1}$,
followed by a strong dip at around 750\,cm$^{-1}$. In addition
there is weaker structure at both lower and higher frequencies.
Carbotte {\it et al.} \cite{carbotte99} argued that the main peak
is due to coupling of charge carriers to a collective bosonic mode
and that it occurs at the frequency $\Delta+\omega_s$, where
$\Delta$ is the maximum gap in the density of states and
$\omega_s$ is the frequency of bosonic mode. They also claimed
that in optimally doped Y123 the spectral weight of the peak
matches that of neutron ($\pi$,$\pi$) resonance and is sufficient
to explain high transition temperature in the cuprates. On the
other hand Abanov {\it et al.} \cite{abanov01} argued that the
main peak due to coupling to collective mode should be at
2$\Delta+\omega_s$. Moreover they argued that the fine structure
at higher frequencies in $\alpha^2$F($\omega$) has physical
significance: the second dip above the main peak should be at
$\omega$=\,4$\Delta$ and the next peak at
$\omega$=\,2$\Delta$+2$\omega_s$.

From Figs.~\ref{fig:num1} and \ref{fig:num2} we conclude that {\it
extreme} caution is required when performing numerical procedures
based on the data smoothed ``by hand". In Fig.~\ref{fig:num2} the
strongest peak at around 480 cm$^{-1}$ is fairly robust, although
its spectral weight does change a few percent. However the other
structures are very dependent on smoothing. The strongest dip
shifts from 780\,cm$^{-1}$ with 11 s.v., to 760\,cm$^{-1}$ with 12
s.v. and 750\,cm$^{-1}$ with 13 s.v. In the data smoothed ``by
hand" it is at 730\,cm$^{-1}$. The spectral weight of the dip also
varies. It was suggested by Abanov {\it et al.} \cite{abanov01}
that the main dip, not the peak, is a better measure of the
frequency 2$\Delta+\omega_s$. However based on our calculations
(Fig.~\ref{fig:num2}) the dip is even more sensitive to smoothing
then the peak. Other peaks and dips do not display any correlation
with the number of s.v., i.e. the level of smoothing.

\section{Model calculations}
\label{model}

\begin{figure}[t]
\vspace*{+0.0cm}%
\centerline{\includegraphics[width=9cm]{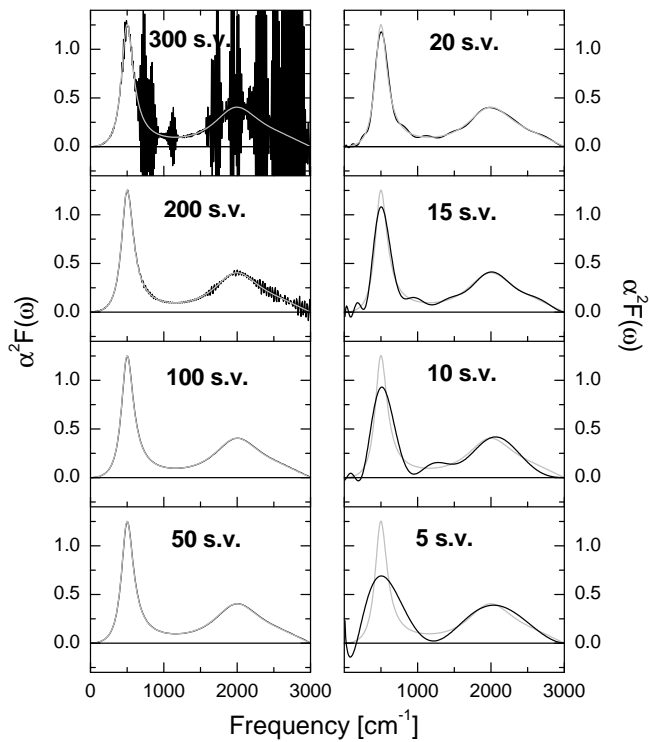}}%
\vspace*{-0.5cm}%
\caption{Model calculations of spectral function with two
Lorentzians (Eq.~\ref{eq:lorz}). The ``exact" solution, with all
300 s.v., does not agree well with the model because small
singular values produce numerical instabilities in the solution.
On the other hand if too few s.v. are kept unphysical negative
regions appear. The model spectral function is recovered with
50--100 s.v.} \label{fig:num3}
\end{figure}

The question we must now try to answer is how many s.v. to keep in
inversion calculations. To address this issue we have performed
calculations based on model spectral function with two
Lorentzians:

\begin{eqnarray}
\alpha^2F(\omega)&=&\frac{\omega_{p,a}^2
\omega^2}{(\omega_a^2-\omega^2)^2+(\gamma_a \omega)^2}
+\nonumber\\&&\frac{\omega_{p,b}^2
\omega^2}{(\omega_b^2-\omega^2)^2+(\gamma_b \omega)^2}, \label{eq:lorz}
\end{eqnarray}
with $\omega_{p,a}^2$=\,50,000\,cm$^{-1}$,
$\omega_a$=\,500\,cm$^{-1}$, $\gamma_a$=\,200\,cm$^{-1}$,
$\omega_{p,a}^2$=\,250,000\,cm$^{-1}$,
$\omega_a$=\,2,000\,cm$^{-1}$ and $\gamma_a$=\,800\,cm$^{-1}$.
This analytic form was chosen to mimic the ``real" spectral
function in cuprates (see for example Fig.~(\ref{fig:positive})
below). From this $\alpha^2$F($\omega$) the scattering rate was
calculated (not shown) using Eq.~(\ref{eq:tau2}) and then the
formalism of inverse theory (Section~\ref{numerical}) was applied.
Figure~\ref{fig:num3} shows the model spectral function (gray
lines) along with the spectral function determined using inverse
theory (black lines). We see that the ``exact" solution (with all
300 s.v.) does not agree well with the model; this is due to
numerical instabilities induced by smallest s.v. As we reduce the
number of s.v. (cut-off the smallest) the agreement improves and
for 100 and 50 s.v. the inversion reproduces the original spectral
function. As we reduce the number of s.v. further the agreement
begins to deteriorate and negative values in $\alpha^2$F($\omega$)
appear again. Obviously these negative values are not real and
simply reflect the fact that too few s.v. do not contain enough
information to reproduce the original data. Note however that even
with very few s.v. the main features of the spectral function are
reproduced, as the main peaks and dips are roughly at correct
frequencies (see for example calculations with 20, 15 and 10
s.v.). Their spectral weights are not reproduced though.

The optimal number of s.v. is always a trade--off between
numerical precision and closeness to the data. Unfortunately,
unlike model calculation shown in Fig.~\ref{fig:num3}, in
calculations with real data those two criteria are not well
separated. Therefore one must be very careful when quantitatively
analyzing the fine structure and their spectral weight in
$\alpha^2$F($\omega$), as different levels and/or methods of
smoothing can cause spurious shifts of the peaks and/or
redistribution of their weights. For example visual inspection of
1/$\tau_{cal}(\omega)$ on the righthand side of
Fig.~\ref{fig:num1} cannot distinguish between different levels of
smoothing [compare 1/$\tau_{cal}(\omega)$ in panels C2, D2 or E2],
however even the smallest differences manifest themselves in the
spectral functions in the left panels.

An advantage of using inverse theory for extracting
$\alpha^2$F($\omega$) is that we can {\it quantify} the smoothing
procedure by specifying the number of s.v. different from zero in
Eq.~(\ref{eq:m2}), thus eliminating arbitrariness related with
smoothing of experimental data ``by hand". This is especially
important when quantitatively comparing results from two different
1/$\tau(\omega)$ curves. Note however that if the data sets have
different signal-to-noise levels, keeping the same number of s.v.
will result in different levels of smoothing. We will encounter
this problem below when we study doping dependence of
$\alpha^2$F($\omega$) in Y123, since available data are from
different sources.

\begin{figure}[t]
\vspace*{+0.0cm}%
\centerline{\includegraphics[width=9cm]{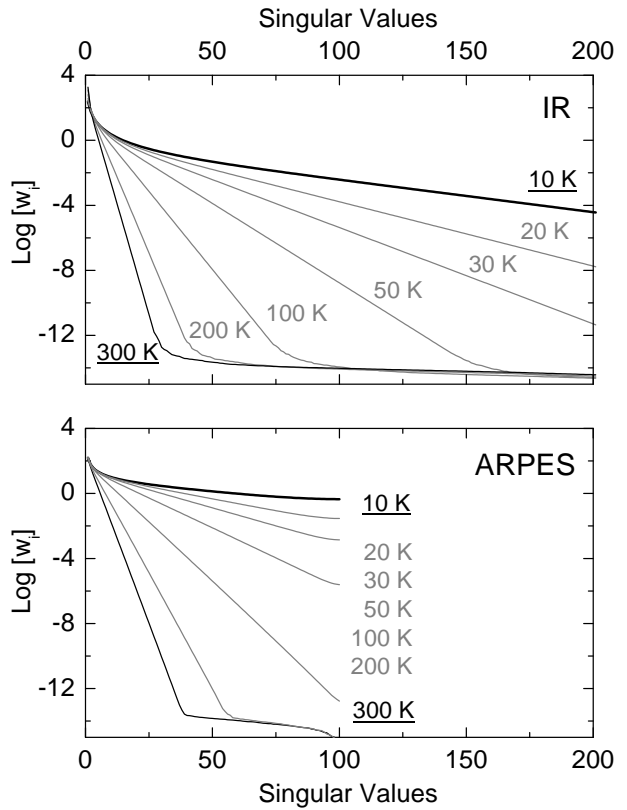}}%
\vspace*{-0.5cm}%
\caption{Singular values at different temperatures. Top panel:
first (biggest) 200 s.v. from IR; bottom panel: all 100 s.v. from
ARPES. When analyzing temperature dependence of the spectral
function ``horizontal cuts" are more appropriate, assuming all
data sets have the same signal-to-noise ratio. On the other hand
for doping dependence studies (at the same temperature) ``vertical
cuts", i.e. the same number of s.v., should produce similar levels
of smoothing.} \label{fig:sv}
\end{figure}

Similar problems arise when analyzing temperature dependence of
the data. Keeping the same number of singular values is again not
the best way to achieve similar levels of smoothing.
Fig.~\ref{fig:sv}A shows the absolute values of first (biggest)
200 s.v. at different temperatures. They drop quickly (note the
$\log$ scale) and such small $w_i$ produce large oscillations in
the solution. To avoid that one cuts--off, i.e. replaces 1/$w_i$
with zeros in Eq.~(\ref{eq:m2}). As Fig.~\ref{fig:sv}A shows s.v.
are also very temperature dependent, and there are different ways
to make the cut. As mentioned above keeping the same number of
s.v. different from zero (``vertical cut") is not a good way, as
that would imply including smaller s.v. at higher temperatures and
therefore higher frequency components into the solution. In such
cases it is better to make ``horizontal cuts", i.e. keep the s.v.
in the same range of absolute values. This implies different
number of s.v. at different temperatures, but the oscillations in
all the solutions should be approximately the same.

\section{Problem of negative values}
\label{negative}

An obvious problem with these (Figs.~\ref{fig:num1} and
\ref{fig:num2}) and previous
(Ref.~\onlinecite{carbotte99,schachinger00,schachinger01,wang01,abanov01,singley01,tu02})
calculations is that they all produce non-physical {\it negative
values} in the spectral function. The latter function is
proportional to boson density of states F($\omega$) and therefore
cannot be negative. The important issue we must address is the
origin of these negative values. As shown in Fig.~\ref{fig:num3}
negative values can appear because of numerical problems: either
because small s.v. produce numerical instabilities, or because too
few s.v. do not contain sufficient information to reproduce the
original data. These negative values are not real and can be
eliminated either by choosing appropriate number of s.v., or by
some other numerical technique, as we will show below.

\begin{figure}[t]
\vspace*{+0.0cm}%
\centerline{\includegraphics[width=9cm]{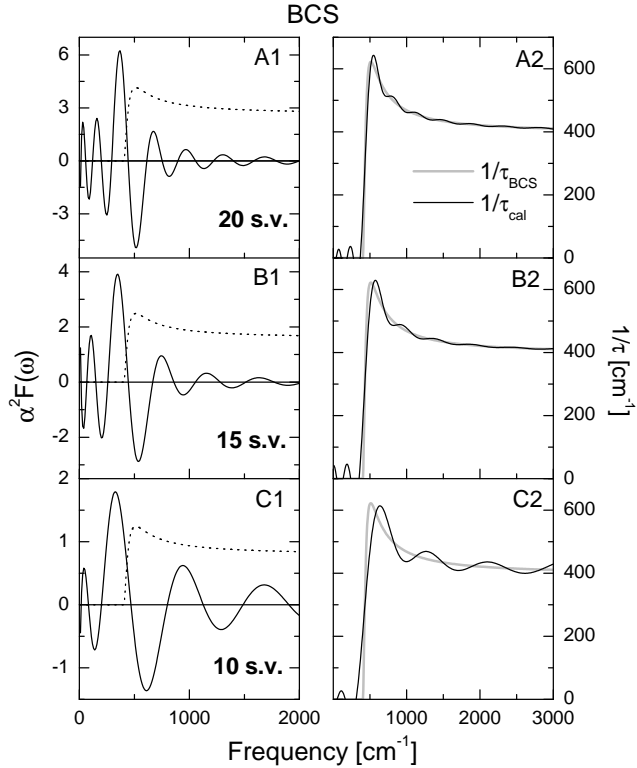}}%
\vspace*{-0.5cm}%
\caption{Model calculations of spectral function from BCS
scattering rate. Left panels display calculated spectral function
$\alpha^2$F($\omega$) and the right panels the BCS scattering rate
(also shown with dotted lines in left panels) and
1/$\tau_{cal}(\omega)$ calculated from the spectral function on
the left. A gap in the density of states produces similar
structure in $\alpha^2$F($\omega$) as does the coupling to a
bosonic mode.} \label{fig:bcs}
\end{figure}

However, negative values can also have a real physical origin, and
they cannot be eliminated by any numerical procedure. Namely, all
the methods we have discussed (Eqs.~(\ref{eq:tau1}), (\ref{eq:w1})
or (\ref{eq:tau2})) were developed for {\it normal} state, but are
frequently used in the (pseudo)gapped state
\cite{abanov01,comm-varma}. In order to illustrate the
insufficiency of these models to account for a (pseudo)gap in the
density of states we have performed inversion calculations on BCS
scattering rate. Fig.~\ref{fig:bcs} shows that scattering rate
(right panels) calculated within BCS with
$\Gamma$=\,2$\Delta$=\,400\,cm$^{-1}$, at T/T$_c$=\,0.1. The
spectral functions calculated with different number of s.v., i.e.
different levels of smoothing are shown in left panels.
Surprisingly they look very similar to those produced by coupling
of carriers to collective bosonic mode (see Figs.~\ref{fig:num1}
and \ref{fig:num2}): there is a strong peak roughly at the
frequency of the gap, followed by a strong dip and fine structure
which is smoothing dependent.

The main issue now is whether one can distinguish between real,
physical negative values arising because of the gap in the density
of states and those arising because of numerical instabilities.
Using inverse theory we can also address this problem. A so called
{\it deterministic constraints} \cite{nr} can be imposed on the
solution during the inversion process. These deterministic
constraints reduce the set of possible solutions from which the
``best" solution will be picked. In the case of spectral function
an obvious constrain is $\alpha^2$F($\omega$)$\geq$\,0 for all
$\omega$. However other constraints are also possible. In fact one
of them, that $\alpha^2$F($\omega$)=\,0 above some cut-off
frequency (3,000\,cm$^{-1}$), was implicitly assumed in all
previous calculations, as the limits in the integral in
Eq.~(\ref{eq:tau2}) run from zero to infinity, and the sum in
Eq.~(\ref{eq:tau4}) runs only up to 3,000\,cm$^{-1}$.

\begin{figure}[t]
\vspace*{-0.6cm}%
\centerline{\includegraphics[width=9cm]{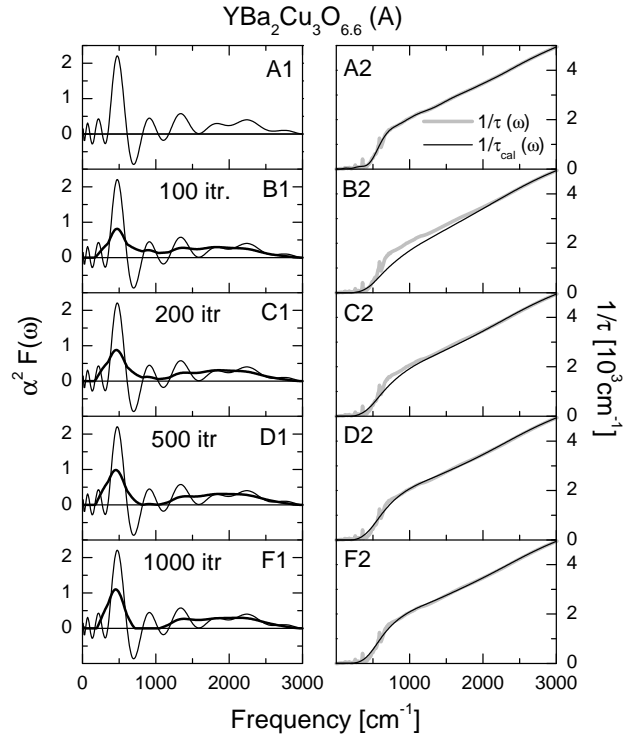}}%
\vspace*{-0.5cm}%
\caption{Spectral function $\alpha^2$F($\omega$) for underdoped
YBa$_2$Cu$_3$O$_{6.6}$ with T$_c$=\,59\,K. A deterministic
constraint $\alpha^2$F($\omega$)$\geq$\,0 is applied iteratively
(Eq.~(\ref{eq:r2})). The top two panels show the initial solution
with 20 s.v. The other four sets of panels display intermediate
solutions for several different levels of iterations.}
\label{fig:positive}
\end{figure}

Numerically one applies the constraints during an iterative
inversion process \cite{nr}. The initial solution $\vec{a}_0$ for
the iteration can be obtained either from Eq.~(\ref{eq:m1}) or
more generally using a so called regularization:

\begin{equation}
K^T \vec{\gamma} = (K^T K+ \delta H) \vec{a}, \label{eq:r1}
\end{equation}
where $H$ is a so called regularization matrix and $\delta$ is a
regularization parameter. For $\delta$=0 (no regularization)
Eq.~(\ref{eq:r1}) reduces to Eq.~(\ref{eq:m1}). Eq.~(\ref{eq:r1})
can also be solved using SVD. Once the initial solution
$\vec{a}_0$ is found, one applies iteration, imposing the
constraint $\alpha^2$F($\omega$)$\geq$\,0 in every step:

\begin{equation}
\vec{a}_{n+1}=P[(I-\beta\delta H)\vec{a}_n+ \beta
K^T(\vec{\gamma}-H \vec{a}_n)], \label{eq:r2}
\end{equation}
where $\beta$ is the iteration parameter and $P$ denotes an
operator that sets all the negative values in the solution to
zero. The results of these calculations for YBa$_2$Cu$_3$O$_{6.6}$
with T$_c$=\,59\,K are shown in Fig.~\ref{fig:positive}. The
initial solution (top panels) was obtained from SVD with 20 s.v.
and no regularization. This solution was then iterated different
number of times: 100 (panel B), 200 (C), 500 (D) and 1000 (F). For
each intermediate solution the scattering rate
1/$\tau_{cal}(\omega)$ (gray lines) was calculated using
Eq.~(\ref{eq:tau2}).

Clearly as the number of iterations increase the agreement between
1/$\tau(\omega)$ and 1/$\tau_{cal}(\omega)$ becomes better, but it
never becomes as good as the one with negative values (top panel).
It appears that the numerical process converges, although very
slowly, to the solution with negative values: some frequency
regions in $\alpha^2$F($\omega$) have simply been cut--off by the
program. The position of the main peak is not affected, but its
intensity has been reduced significantly. We also emphasize that
the structure in the spectral function at
$\omega>$\,1,000\,cm$^{-1}$ is essential for obtaining linear
frequency dependence of 1/$\tau(\omega)$ up to very high
frequencies. We will return to this important issue in
Section~\ref{comparison} below.

\begin{figure}[t]
\vspace*{-1.0cm}%
\centerline{\includegraphics[width=9cm]{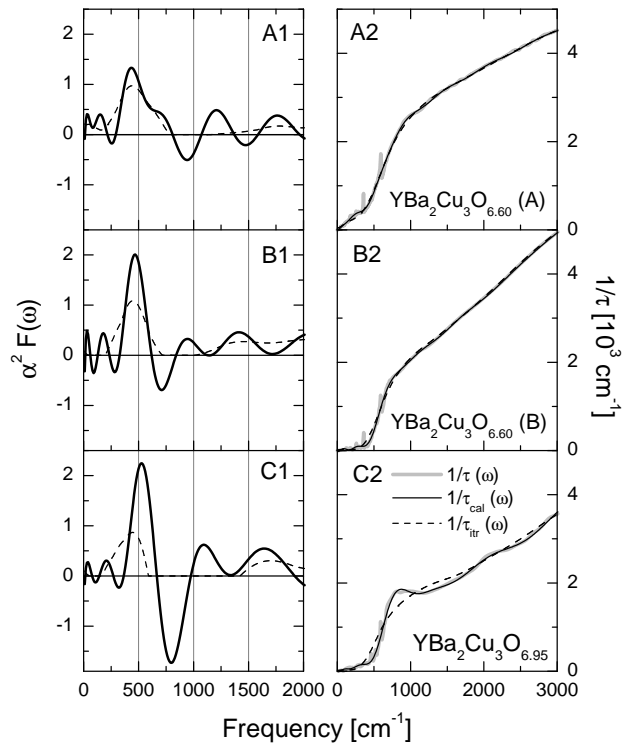}}%
\vspace*{-0.5cm}%
\caption{Doping dependence of spectral function
$\alpha^2$F($\omega$) for (A) YBa$_2$Cu$_3$O$_{6.60}$ with
T$_c$\,=\,57\,K (Ref.~\protect\onlinecite{homes99}), (B)
YBa$_2$Cu$_3$O$_{6.60}$ with T$_c$\,=\,59\,K
(Ref.~\protect\onlinecite{carbotte99}) and (C)
YBa$_2$Cu$_3$O$_{6.95}$ with T$_c$\,=\,91\,K
(Ref.~\protect\onlinecite{homes99}). All curves are for the lowest
measured temperature T$\approx$\,10\,K. ``Vertical cuts", i.e. the
same number of singular values (15), were made for all three data
sets. Left panels display 1/$\tau(\omega)$ data along with
1/$\tau_{cal}(\omega)$. Dashed lines are the results of iterative
calculations. Relatively good fits without negative values in
$\alpha^2$F($\omega$) can be obtained for both
YBa$_2$Cu$_3$O$_{6.60}$ samples, but not for
YBa$_2$Cu$_3$O$_{6.95}$.} \label{fig:ybco}
\end{figure}

Therefore in the case of YBa$_2$Cu$_3$O$_{6.60}$ we have been able
to eliminate negative values and at least in principle obtain
$\alpha^2$F($\omega$) which is always positive. This indicates
that the structure in the spectral function is (predominantly) due
to coupling to bosonic mode and not the gap in the density of
states. On the other hand, we have not been able to obtain good
BCS scattering rate without negative values in spectral function
(not shown). This is not unexpected as the form of spectral
function is entirely due to a gap in the density of states (no
bosonic mode), which Eqs.~(\ref{eq:tau1}) and (\ref{eq:tau2}) do
not take into account. We have encountered similar situation in
some cuprates. Fig.~\ref{fig:ybco} displays inversion calculations
for several Y123 samples with different doping levels and/or
T$_c$: YBa$_2$Cu$_3$O$_{6.60}$ with T$_c$\,=\,57\,K
(Ref.~\onlinecite{homes99}), YBa$_2$Cu$_3$O$_{6.60}$ with
T$_c$\,=\,59\,K (Ref.~\onlinecite{carbotte99}) and
YBa$_2$Cu$_3$O$_{6.95}$ with T$_c$\,=\,91\,K
(Ref.~\onlinecite{homes99}). All calculations are for T=\,10\,K,
with a fixed number of 15 s.v. As can be seen from
Fig.~\ref{fig:ybco} the peak systematically shifts to higher
energies as doping and T$_c$ increase: 430\,cm$^{-1}$ in x=6.6
with T$_c$=\,57\,K, 480\,cm$^{-1}$ in the second x=6.6 with
T$_c$=\,59\,K and 520\,cm$^{-1}$ in x=6.95. For both 6.6 samples
we have been able to obtain relatively good inversions (dashed
lines) without negative values in the spectral function. That is
not the case for 6.95 sample where without negative values the
inversion fails badly (see dashed line in the bottom--right
panel). This indicates that the form of scattering rate is
probably a combination of coupling to collective mode and a gap in
the density of states, as pointed out by Timusk \cite{timusk03}.

\section{Electron-boson coupling vs. energy gap}
\label{sc}

As demonstrated in previous sections similar shapes of
$\alpha^2F(\omega)$ are produced by coupling to bosonic mode and a
gap in the density of states when equations for the normal state
(Eq.~(\ref{eq:tau1}) or (\ref{eq:tau2})) are used. It is essential
to discriminate these two contribution because they usually appear
together. To address this problem we have to apply Allen's formula
for the scattering rate in the superconducting state
\cite{allen71}:

\begin{equation}
\frac{1}{\tau(\omega)}=\frac{2\pi}{\omega}\int_{0}^{\omega-2\Delta}
d\Omega (\omega-\Omega) \alpha^2F(\Omega) E \Big[ \sqrt{1-\frac{4
\Delta^2}{(\omega-\Omega)^2} } \Big] \label{eq:bcs-tau}
\end{equation}
In this equation $E(x)$ is the complete elliptic integral of the
second kind and $\Delta$ is a gap in the density of states. For
$\Delta$=0 Eq.~(\ref{eq:bcs-tau}) reduces to Eq.~(\ref{eq:tau1})
for the normal state. Numerically Eq.~(\ref{eq:bcs-tau}) is again
Fredholm integral equation of the second kind and the same
numerical procedure for its solution can be used.

\begin{figure}[t]
\vspace*{+0.0cm}%
\centerline{\includegraphics[width=9cm]{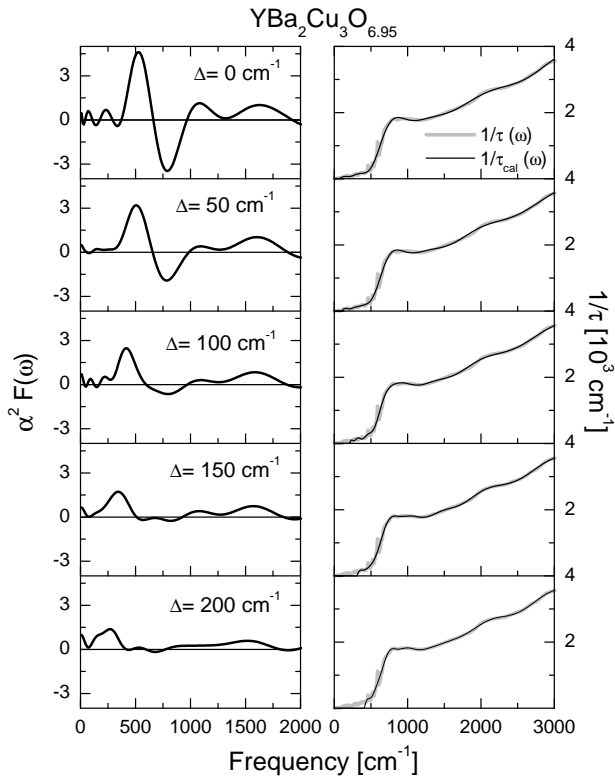}}%
\vspace*{-0.5cm}%
\caption{Spectral function $\alpha^2$F($\omega$) for optimally
doped YBa$_2$Cu$_3$O$_{6.95}$ calculated from
Eq.~(\ref{eq:bcs-tau}) for the scattering rate in the
superconducting state. Different values of the gap
$\Delta$=0--200\,cm$^{-1}$ were used in the calculations. For
$\Delta$=0 there is a pronounced deep following the main peak.
However when finite values of the gap are introduced the negative
deep gradually disappears and the main peak shifts to lower
energies.} \label{fig:alpha-gap}
\end{figure}

We have performed inversion of the data for optimally doped
YBa$_2$Cu$_3$O$_{6.95}$ using Eq.~(\ref{eq:bcs-tau}).
Fig.~\ref{fig:alpha-gap} shows inversion calculations for
different values of the gap $\Delta$. The top panels display
calculations with $\Delta$=0, which is equivalent to previous
calculations using Eq.~(\ref{eq:tau2}) (Fig.~\ref{fig:ybco}). As
we already discussed it, the spectrum is dominated by a pronounced
peak, followed by a large negative deep. Unlike
YBa$_2$Cu$_3$O$_{6.6}$ for which this negative deep can, at least
in principle, be eliminated, the deep in YBa$_2$Cu$_3$O$_{6.95}$
cannot be eliminated (Fig.~\ref{fig:ybco}) and in the previous
section we suggested that that is because of the gap. Indeed when
finite values of the gap are used in Eq.~(\ref{eq:bcs-tau}) this
negative deep following the main peak is strongly suppressed;
calculated spectral function positive for almost all frequencies
(Fig.~\ref{fig:alpha-gap}).

\begin{figure}[t]
\vspace*{+2.5cm}%
\centerline{\includegraphics[width=9cm]{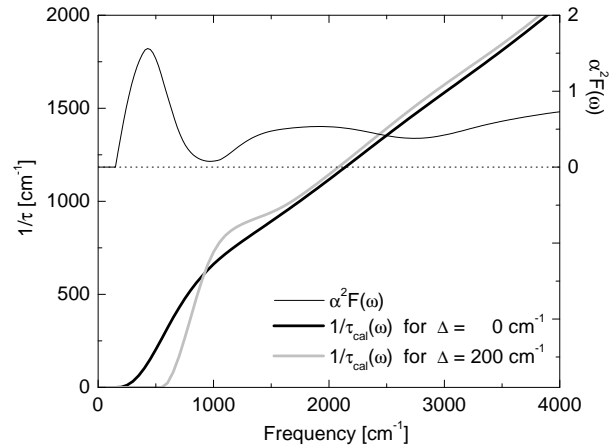}}%
\vspace*{-2.0cm}%
\caption{Model spectral function $\alpha^2$F($\omega$) (thin line)
is used to calculate the scattering rate 1/$\tau_{cal}(\omega)$
from Eq.~(\ref{eq:bcs-tau}). For $\Delta$=0 calculated scattering
rate resembles 1/$\tau(\omega)$ of underdoped
YBa$_2$Cu$_3$O$_{6.60}$ (Fig.~\ref{fig:ybco}). However for finite
values of the gap calculated scattering rate resembles
1/$\tau(\omega)$ of optimally doped YBa$_2$Cu$_3$O$_{6.95}$: there
is an {\it overshoot} following the suppressed region
(Fig.~\ref{fig:ybco}).} \label{fig:model-gap}
\end{figure}

The problem with Eq.~(\ref{eq:bcs-tau}) is that it is based on
s-wave energy gap at T=\,0\,K. These two assumptions imply that
the scattering rate must be zero below 2$\Delta$, which is never
the case with cuprates because of the d-wave gap and because the
data was taken at finite temperature. In spite of this,
Eq.~(\ref{eq:bcs-tau}) is useful because it can provide some
insight into charge dynamics in cuprates. Fig.~\ref{fig:model-gap}
displays calculations of scattering rate based on model spectral
function $\alpha^2$F($\omega$) shown with thin line. When the gap
is zero the data qualitatively looks like underdoped
YBa$_2$Cu$_3$O$_{6.6}$: at higher frequencies it is linear and it
is suppressed below certain energy (black line). However for the
finite values of the gap ($\Delta$=200\,cm$^{-1}$) the data looks
more like optimal YBa$_2$Cu$_3$O$_{6.95}$: there is overshoot just
above the suppressed region (gray line).

Based on these model calculations it appears that the response of
YBCO on the underdopd side is dominated by coupling to bosonic
mode, whereas at optimal doping the gap plays more prominent role.
Indeed recent ARPES and tunneling measurements have shown that the
Fermi surface of cupares is continuously destroyed with
underdoping \cite{norman98,mcelroy04}. On the underdoped side
antinodal states do not exist (they are incoherent) and the IR
response is dominated by nodal states which are coherent and not
gaped. On the other hand the IR response at optimal doping is more
complicated, because both antinodal (gaped) and nodal (not gaped)
states are coherent and contribute to the IR response.

\section{Electron--boson spectral function of Bi2212}
\label{bi2212}

\begin{figure}[t]
\vspace*{+0.0cm}%
\centerline{\includegraphics[width=9cm]{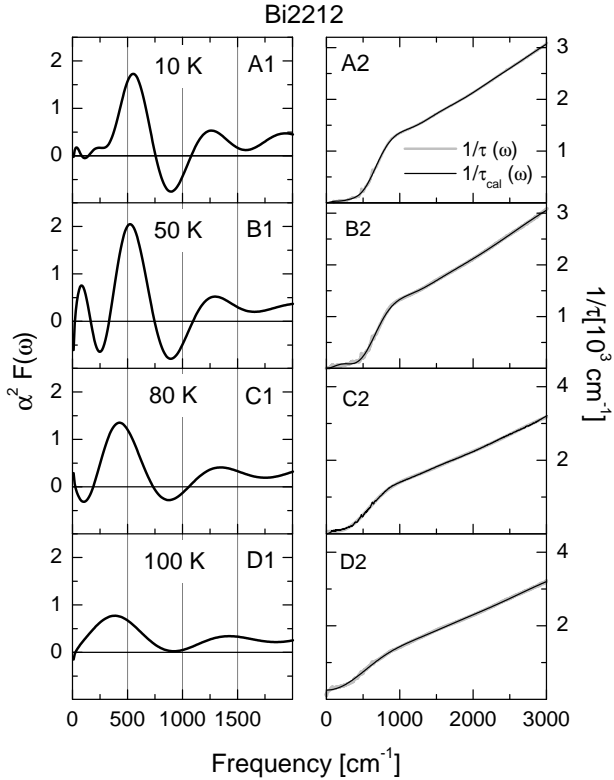}}%
\vspace*{-0.5cm}%
\caption{Temperature dependence of the spectral function
$\alpha^2$F($\omega$) for optimally doped
Bi$_2$Sr$_2$CaCu$_2$O$_{8-\delta}$ with T$_c$=\,91\,K. As
temperature increases the main peak shifts to lower energies and
looses intensity, but seems to persist even above T$_c$.}
\label{fig:bi2212}
\end{figure}

In this section we analyze the temperature dependence of the
spectral function for optimally doped Bi2212 with T$_c$=91\,K. The
same data set has been analyzed before \cite{tu02} using
Eq.~(\ref{eq:w1}). The calculated spectra (Fig.~\ref{fig:bi2212})
look qualitatively similar to those obtained on Y123
(Fig.~\ref{fig:num2}), with a strong peak in the far-IR range
followed by a dip, and high frequency contribution that extends up
to several thousand cm$^{-1}$. To achieve similar levels of
smoothing at different temperatures ``horizontal cut" has been
made (Fig.~\ref{fig:sv}A).

In the normal state at T=100\,K we identify a peak at
$\approx$\,400\,cm$^{-1}$ (50\,meV). Note that the peak is at
somewhat lower energy then in Ref.~\onlinecite{tu02}, which can be
traced back to the use of Eq.~(\ref{eq:w1}), which is strictly
speaking valid only at T=0\,K. We also note that the peak is
observed above T$_c$, unlike the ($\pi$, $\pi$) resonance detected
in INS only in the superconducting state \cite{fong99}.

As temperature decreases below T$_c$ the peak shifts to higher
energies: 430\,cm$^{-1}$ at 80\,K, 520\,cm$^{-1}$ at 50\,K and
560\,cm$^{-1}$ at 10\,K. At the lowest temperature the spectral
function is almost identical to previously reported \cite{tu02},
which confirms that at 10\,K Eqs.~(\ref{eq:w1}) and
(\ref{eq:tau2}) are equivalent. According to theoretical
considerations \cite{carbotte99,abanov01} in the superconducting
state the peak should be off-set from the resonance frequency of
the ($\pi$, $\pi$) peak ($\omega_s\simeq$\,43\,meV) by one or two
gap values ($\Delta$=34\,meV, Ref.~\onlinecite{rubhausen98}). At
10\,K the peak is at 70\,meV, somewhat lower than
$\Delta+\omega_s$=\,77\,meV (Ref.~\onlinecite{carbotte99}) and
significantly lower than 2$\Delta+\omega_s$=\,111\,meV
(Ref.~\onlinecite{abanov01}). This result is in contrast with
optimally doped Y123 where the IR peak at 66\,meV
(Fig.~\ref{fig:ybco}) is in relatively good agreement with
$\Delta+\omega_s$=27\,meV+41\,meV=\,68\,meV
Ref.~\onlinecite{carbotte99}.

\section{Inversion of ARPES data}
\label{arpes}

Recently it has been argued based on ARPES data
\cite{bogdanov00,lanzara01,shen03,zhou03} that in cuprates
electrons are strongly coupled to phonons and that such strong
coupling might be responsible for high T$_c$. In light of these
suggestions there have been several attempts to determine
$\alpha^2$F($\omega$) from ARPES data
\cite{verga02,schachinger03,shi04}. Inversion by Verga {\it et
al.} \cite{verga02} was based on the imaginary part of the
self-energy $\Sigma_2(\omega)$, obtained from the real part
$\Sigma_1(\omega)$ through Kramers--Kronig transformation. The
spectral function was then calculated by differentiation of
$\Sigma_2(\omega)$, a procedure which necessarily requires
smoothing ``by hand". On the other hand Schachinger {\it et al.}
\cite{schachinger03} have modeled the spectral function with
analytical functions and then used these models to simultaneously
fit both the IR and ARPES spectra. Maximum Entropy Method (MEM)
has recently been used to invert ARPES data and obtain
$\alpha^2$F($\omega$) for beryllium surface Be(10$\overline{1}$0)
(Ref.~\onlinecite{shi04}) and LSCO (Ref.~\onlinecite{zhou04}).
Here we apply the same inversion method we used for IR to ARPES.
The procedure of extracting $\alpha^2$F($\omega$) is based on
standard expression for the real part of quasiparticle self-energy
$\Sigma_1(\omega)$ (Ref.~\onlinecite{allen82}):

\begin{eqnarray}
\Sigma_1(\omega)&=&\int_{0}^{\infty}d\Omega \alpha^2F(\Omega)\cdot
\nonumber \\ && \Re \Big [\Psi \Big(
\frac{1}{2}+i\frac{\Omega-\omega}{2\pi T} \Big )- \Psi \Big(
\frac{1}{2}-i\frac{\Omega+\omega}{2\pi T} \Big) \Big] ,
\label{eq:arpes2}
\end{eqnarray}
where $\Psi(x)$ is digamma function. The real part of the
self-energy $\Sigma_1(E_k)$ can be obtained from ARPES data as
\cite{allen82}:

\begin{equation}
\Sigma_1(E_k)=E_k-\epsilon_k, \label{eq:arpes3}
\end{equation}
where E$_k$ is the renormalized dispersion measured in ARPES
experiments and $\epsilon_k$ is the bare electron dispersion. As
the latter function is not independently known, a common procedure
when using Eq.~(\ref{eq:arpes3}) is to assume a linear bare
dispersion ($\epsilon_k\sim k$) and no renormalization at higher
energies, i.e. E$_k$=$\epsilon_k$ above $\approx$\,250\,meV.
Expression (\ref{eq:arpes2}) is again a Fredholm integral equation
of the first kind and the same numerical technique described in
Section~\ref{numerical} can be used for its solution. Similar to
IR, ``by hand" smoothing of the data is not needed, as SVD
procedure will allow us to smooth the solution by reducing the
number of non-zero s.v. Since the resolution of ARPES data is
poorer than IR, in all calculations we used vectors and matrices
with dimensions 100 instead of 300.

\begin{figure}[t]
\vspace*{+0.0cm}%
\centerline{\includegraphics[width=9cm]{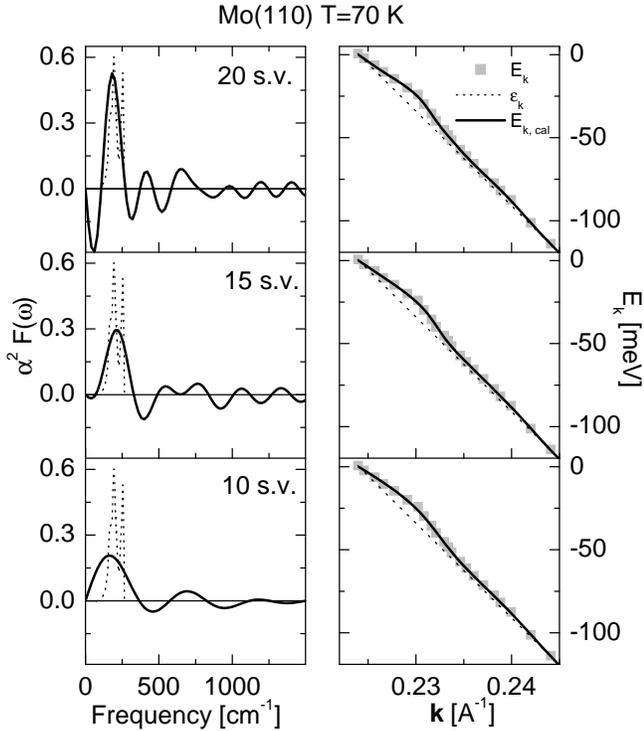}}%
\vspace*{-0.5cm}%
\caption{Spectral function $\alpha^2$F($\omega$) of Mo(110)
surface at 70\,K extracted from ARPES data \protect\cite{valla99}.
Three different levels of smoothing are shown with 10, 15 and
20\,s.v. Dotted lines in left panels represent theoretical
spectral function \protect\cite{savrasov96}.} \label{fig:mo}
\end{figure}

As an example of this procedure in Figure~\ref{fig:mo} we first
present spectral function $\alpha^2$F($\omega$) calculated from
ARPES data for molybdenum surface Mo(110)
(Ref.~\onlinecite{valla99}). As before, left panels show the
calculated spectral function and right panels measured ARPES
dispersion E$_k$ and dispersion calculated from
Eq.~(\ref{eq:arpes2}) E$_{k,cal}$ using the corresponding spectral
function on the left. The spectral function has a characteristic
shape, with a strong peak at around 200\,cm$^{-1}$ and weaker
structure at both higher and lower frequencies. Similar to IR,
position of the main peak is fairly robust against smoothing, but
weaker peaks and dips are not. The dashed lines in the left-hand
panels represent $\alpha^2$F($\omega$) calculated based on band
structure \cite{savrasov96}. Low data resolution and loss of
information during the inversion do not allow us to resolve the
fine structure in $\alpha^2$F($\omega$) that has been predicted
numerically \cite{savrasov96}. At higher energies
($\omega\gtrsim$\,400\,cm$^{-1}$) the spectral function is
effectively zero, in accord with band structure calculations.

These relatively simple calculations for molybdenum surface
Mo(110) have uncovered the limitations of inversion of ARPES data.
Fine details of the spectral function, especially narrow peaks,
cannot be resolved as they are convoluted in the experimental data
(Eq~(\ref{eq:arpes2})). Maximum information that can be obtained
is the {\it frequency region} where there is significant
contribution to $\alpha^2$F($\omega$). It has recently been
claimed based on MEM inversion of ARPES data that the sharp peaks
identified in $\alpha^2$F($\omega$) spectra are due to specific
phonon modes \cite{shi04,zhou04}. Based on our calculations we
speculate that it is unlikely that such fine details of the
spectra could be resolved by any inversion procedure.

\begin{figure}[t]
\vspace*{+2.5cm}%
\centerline{\includegraphics[width=9cm]{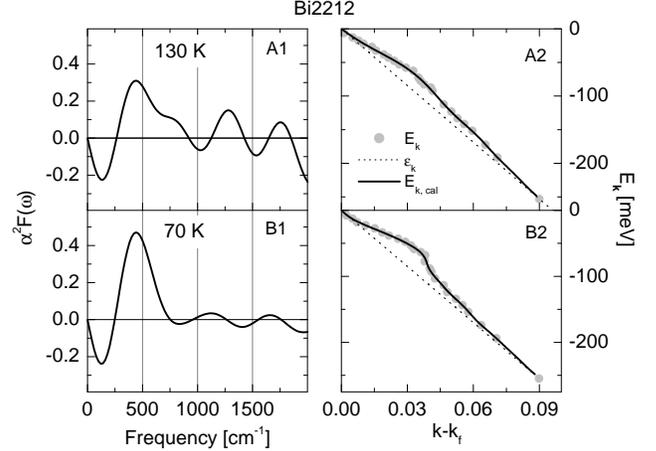}}%
\vspace*{-2.0cm}%
\caption{Temperature dependence of spectral function
$\alpha^2$F($\omega$) of optimally doped Bi2212 extracted from
ARPES data (along nodal direction) using inverse theory. Left
panels show $\alpha^2$F($\omega$) spectra calculated from
Eq.~(\ref{eq:arpes2}), and right panels ARPES quasiparticle
dispersion E$_k$ (gray symbols) and calculated dispersion
E$_{k,cal}$ (full lines) using corresponding spectral function on
the left. Also shown with dashed lines are bare quasiparticle
dispersions $\epsilon_k$ used to calculate $\Sigma_1(\omega)$
(Eq.~(\ref{eq:arpes3})).} \label{fig:arpes}
\end{figure}

Figure~\ref{fig:arpes} presents the data for optimally doped
Bi2212 (T$_c$=\,91\,K) at 130\,K and 70\,K taken along nodal
direction. Similar to IR calculations in Fig.~\ref{fig:bi2212}, to
achieve approximately the same level of smoothing different number
of s.v. values were kept in calculations at different
temperatures: 8 (out of 100) at 130\,K and 10 at 70\,K. Within the
error bars the main peak does not shift with temperature: it is at
440\,cm$^{-1}$ at both 130\,K and 70\,K. However the peak does
narrow and gains strength at 70\,K (Ref.~\onlinecite{arpes-sv}).
Below 70\,K ARPES dispersion displays almost no temperature
dependence. Note also that unlike IR, there seems to be less
problems with negative values in ARPES $\alpha^2$F($\omega$)
calculations. In particular there is no pronounced dip following
the main peak, which might be related to the fact that the APRES
scans were taken along ($\pi$, $\pi$) direction where the
magnitude of the gap goes to zero. Another important difference
compared with IR is that there is no high frequency component in
ARPES: the whole contribution to $\alpha^2$F($\omega$) is
concentrated at $\omega\lesssim$\,750\,cm$^{-1}$.

\section{IR--ARPES comparison}
\label{comparison}

In the previous section the inversion calculations have uncovered
several important differences between the spectral function
extracted from IR and ARPES. In all ARPES calculations the strong
dip following the main peak was absent, which we suggested was due
to the absence of the gap along the ($\pi$, $\pi$) symmetry
direction. More importantly, there was no high frequency
contribution extending up to several thousand cm$^{-1}$ in any
ARPES calculations. In this section we will make an explicit
comparison between IR and ARPES spectral functions and discuss
their similarities and differences.

First it should be emphasized that ARPES $\alpha^2$F($\omega$)
from Eq.~(\ref{eq:arpes2}) is not the same as the IR from
Eqs.~(\ref{eq:tau1}) and (\ref{eq:tau2})
\cite{allen82,schachinger03}. ARPES is a momentum resolving
technique, whereas IR averages over the Brillouin zone. More
importantly ARPES probes the equilibrium $\alpha^2$F($\omega$)
(single-particle property), whereas IR measures transport
$\alpha_{tr}^2$F($\omega$) (two-particle property) \cite{allen82}.
Recently Schachinger, {\it et al.} discussed the difference
\cite{schachinger03} and suggested that in the simplest case these
two functions might differ only by a numerical factor of 2--3.
Therefore it would be very instructive to directly compare the
spectral functions extracted from IR and ARPES. However technical
reasons make this comparison difficult. Present resolution of
ARPES data of $\approx$\,10\,meV is at least one order of
magnitude less than IR (typically $\lesssim$\,1\,meV in the
frequency range of interest). This large discrepancy in resolution
requires different levels of smoothing, which can affect the
solution. Therefore we cautiously compare calculated
$\alpha^2$F($\omega$)'s, relying only on robust features, as
discussed in the previous sections.

\begin{figure}[t]
\vspace*{+2.0cm}%
\centerline{\includegraphics[width=9cm]{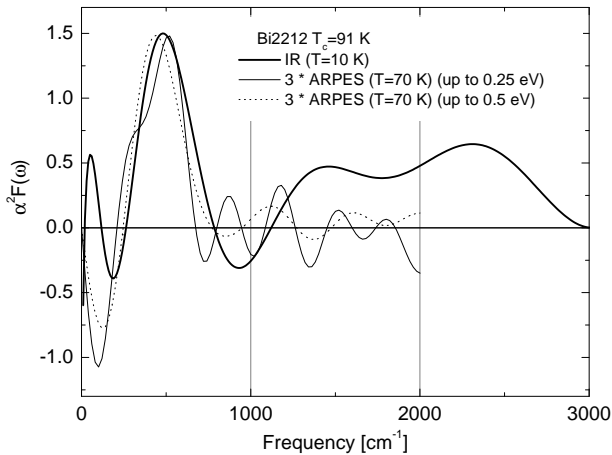}}%
\vspace*{-2.0cm}%
\caption{Comparison of spectral functions $\alpha^2$F($\omega$)
extracted from IR and ARPES data for optimally doped Bi2212 with
T$_c$=\,91\,K. Note that $\alpha^2$F($\omega$) from ARPES is
multiplied by a factor of three. Dashed line is an ARPES
calculation with a different bare dispersion (with renormalization
effects up to 0.5\,eV).} \label{fig:comp}
\end{figure}

The most complete comparison can be made for optimally doped
Bi2212, for which high quality IR \cite{tu02} and ARPES
\cite{fedorov99} data sets exist, all obtained on the samples from
the same batch \cite{gu90}. Although data at different
temperatures are available, we believe that will not change the
main results and conclusions in any significant way, as ARPES data
display little temperature dependence below 70\,K
(Ref~\onlinecite{fedorov99}). Fig.~\ref{fig:comp} shows
$\alpha^2$F($\omega$) from both IR and ARPES for optimally doped
Bi2212 with T$_c$=\,91\,K. The $\alpha^2$F($\omega$) from ARPES is
multiplied by a factor of 3. The agreement between the positions
of the main peak in both data sets ($\approx$\,500\,cm$^{-1}$) is
very good. This agreement is actually surprising and unexpected.
As discussed in Sections~\ref{example} and \ref{bi2212} the main
peak in IR spectral function should be off-set from the frequency
of the neutron peak $\omega_s$ by either one \cite{carbotte99} or
two \cite{abanov01} gap values $\Delta$. On the other hand in
ARPES data the gap should not play a role: the data were taken
along the nodal directions where the gap is zero. Therefore almost
perfect agreement between the positions of IR and ARPES peaks (and
disagreement with INS peak - see Section~\ref{bi2212}) in
optimally doped Bi2212 is puzzling and calls for further
theoretical studies.

\begin{figure}[t]
\vspace*{+0.0cm}%
\centerline{\includegraphics[width=9cm]{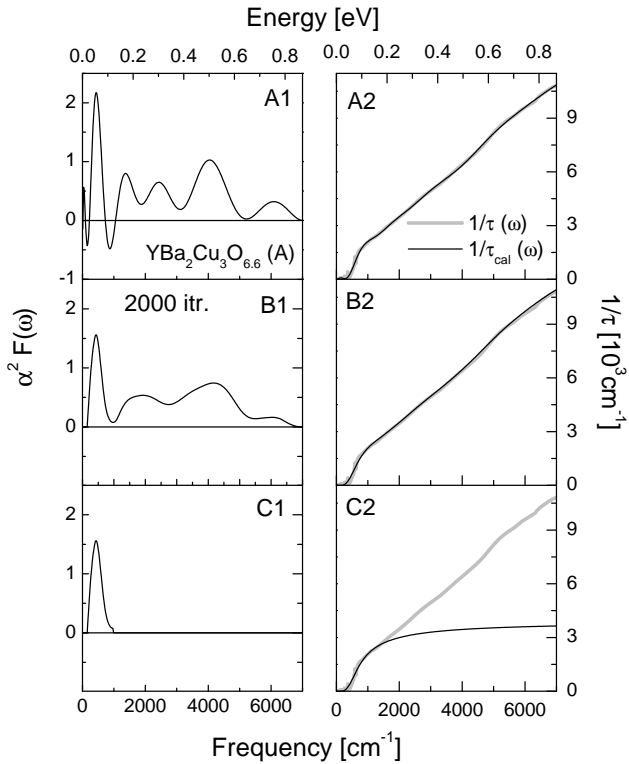}}%
\vspace*{-0.5cm}%
\caption{Spectral function $\alpha^2$F($\omega$) extracted from IR
data for underdoped YBa$_2$Cu$_3$O$_{6.6}$ with T$_c$=\,59\,K and
calculated up to $\approx$\,0.85\,eV. Top panels display
calculations with negative values (Section~\ref{numerical}).
Middle panels are iterative calculations with 2,000 iterations
(Section~\ref{negative}). In both cases significant contribution
to $\alpha^2$F($\omega$) persist up to very high frequencies.
Bottom panels display the result of calculation of
1/$\tau_{cal}(\omega)$ with high frequency contribution to
$\alpha^2$F($\omega$) cut-off (above 1,000\,cm$^{-1}$). In this
case the calculated scattering rate tends to saturate at higher
energies.} \label{fig:long}
\end{figure}

Another important difference between IR and ARPES is the
contribution in IR that extends up to very high energies. There is
no such contribution in any ARPES data we have available
(Section~\ref{arpes}). Therefore based on ARPES data alone one can
argue that the observed contribution to $\alpha^2$F($\omega$) is
either due to phonons or spin fluctuations. On the other hand the
high frequency component in always present in IR and is necessary
to keep 1/$\tau(\omega)$ increasing, approximately linearly with
$\omega$. Figure~\ref{fig:long} displays calculations of
$\alpha^2$F($\omega$) for YBa$_2$Cu$_3$O$_{6.6}$ with
T$_c$=\,59\,K up to almost 1\,eV (Ref.~\onlinecite{carbotte99}).
Both inversion with negative values (top panels) and iterative
calculations with positive values (middle panels) result in
spectral function with significant contributions up to
$\approx$\,0.85\,eV. If this contribution is cut off, for example
at 1,000\,cm$^{-1}$ (bottom panels), calculated scattering rate
deviates strongly from experimental data, as
1/$\tau_{cal}(\omega)$ tends to saturate above
$\sim$\,2,000\,cm$^{-1}$. This result argues against phonons as
the origin of the structure in $\alpha^2$F($\omega$), as phonon
spectrum cannot extend up to such high frequencies. However phonon
contribution below $\approx$\,1,000\,cm$^{-1}$ cannot be ruled out.

The absence of high-frequency contribution in ARPES is puzzling
and seems to indicate that the difference between IR and ARPES
spectral function might be more than just a numerical prefactor.
On the other hand it may also signal intrinsic problems with our
procedure of extracting $\Sigma_1(E_k)$ from ARPES dispersion
\cite{kordyuk04}. As mentioned in Section~\ref{arpes}, bare electron
dispersion $\epsilon_k$ is not known and some assumptions must be
made before Eq.~(\ref{eq:arpes3}) can be used. The most common
assumptions are: 1) linear bare dispersion $\epsilon_k$ and 2) no
renormalization above certain cut-off frequency. We have employed
these assumptions in all our calculations, with a cut-off of
typically $\approx$\,250\,meV. The use of both of these
assumptions in highly unconventional systems like cuprates is
questionable and requires further theoretical treatment
\cite{kordyuk04}.

In order to check the effect upper cut-off energy has on the
solution we have performed $\alpha^2$F($\omega$) inversion for
optimally doped Bi2212 (Fig.~\ref{fig:arpes}) assuming that the
renormalization persist up to 0.5\,eV, instead of 0.25\,eV.
Fig.~\ref{fig:comp} also shows this new calculation with dashed
line and obviously there is very little difference: the main peak
is in good agreement and there is no significant contribution
above $\approx$\,800\,cm$^{-1}$, even though the renormalization
extends up to 0.5\,eV. We speculate that in order to obtain
spectral function similar to IR, either the renormalization must
persist up to several eV or some more sophisticated form of the bare
dispersion $\epsilon_k$ must be used \cite{kordyuk04}.

\section{Summary and Outlook}
\label{conclusions}

A new numerical procedure of extracting electron--boson spectral
function from IR and ARPES data based on inverse theory has been
presented. The new method eliminates the need for differentiation
and smoothing ``by hand". However we also showed that the
information is convoluted and fine details of
$\alpha^2$F($\omega$) cannot be extracted, no matter what
numerical technique one uses. This especially holds for ARPES,
whose current data resolution is particularly poor compared to IR.

Using this new procedure we have extracted $\alpha^2$F($\omega$)
from IR and/or ARPES data in a series of Y123 and Bi2212 samples.
The calculations have uncovered several important differences
between IR and ARPES spectral functions. All IR spectral functions
contain, in addition to a strong peak at low frequencies
($\omega\lesssim$\,500\,cm$^{-1}$), contributions that extend up
to very high energies (typically several thousand cm$^{-1}$). On
the other hand none of ARPES spectral functions display such high
energy contribution. Therefore we concluded that based on ARPES
results one cannot distinguish between phonon and magnetic
scenarios, as the main peak in $\alpha^2$F($\omega$) can have
either (or both) magnetic or phonon components. However in all IR
results the observed high frequency contribution extends to much
higher than typical phonon frequencies, the result which argues
against phonon mechanism.

Finally, the observed differences between IR and ARPES have
prompted us to speculate that $\alpha^2$F($\omega$) from these two
experimental techniques might contain qualitatively different
information. Alternatively, we suggest that the whole concept of
coupling of charge carriers to collective boson modes in the
cuprates needs to be revised.

%
% Acknowledgements...
%
\begin{acknowledgments}
We thank J.P.~Carbotte and A.V.~Chubukov for useful discussions.
Special thanks to M.~Trajkovic for assistance with numerical
solution of integral equations. The research supported by the U.S.
Department of Energy under Contract No. DE-AC02-98CH10886.
\end{acknowledgments}

%
% References
%


\begin{thebibliography}{99}
\bibitem{inverse-comm} Attempts of using inverse theory have been
reported \cite{timusk76,shulga01,shi04}.
\bibitem{comm-allen} Inverse theory can also be applied to
Allen's formula Eq.~(\ref{eq:tau1}), but obviously the result will
only be valid at T=\,0\,K.
\bibitem{quadratic-comm} In general, matrix K can be rectangular,
i.e. vectors $\vec{a}$ and $\vec{\gamma}$ do not have the same
size \cite{nr}. For simplicity in all our calculations we used
quadratic matrices.
\bibitem{comm-varma} An exact numerical treatment has recently
been proposed \cite{varma03}, but this is beyond the scope of our
discussion.
\bibitem{arpes-sv} Fig.~\ref{fig:sv}B shows the absolute values
of s.v. for ARPES data. Compared with corresponding IR, ARPES
curves are flatter and are less sensitive to temperature. This
might be the reason the same features are less sensitive to
smoothing and are more pronounced at higher temperature in ARPES
then in IR.
\end{thebibliography}
\end{document}